\documentclass[amsmath,amssymb,aps,pra,superscriptaddress,twocolumn
]{revtex4-2}

\usepackage{amssymb}
\usepackage{graphicx}% Include figure files
\usepackage{dcolumn}% Align table columns on decimal point
\usepackage{bm}% bold math
\usepackage{amsmath}
\usepackage[colorlinks,urlcolor=cyan,citecolor=blue,linkcolor=magenta]{hyperref}

\begin{document}
\title{Collective excitations in two-dimensional harmonically trapped quantum droplets}

\author{Yifan Fei}
\affiliation{Department of Physics and Key Laboratory of Optical Field Manipulation of Zhejiang Province, Zhejiang Sci-Tech University, Hangzhou 310018, China}

\author{Xucong Du}
\affiliation{Department of Physics and Key Laboratory of Optical Field Manipulation of Zhejiang Province, Zhejiang Sci-Tech University, Hangzhou 310018, China}

\author{Xiao-Long Chen}
\email{xiaolongchen@zstu.edu.cn}
\affiliation{Department of Physics and Key Laboratory of Optical Field Manipulation of Zhejiang Province, Zhejiang Sci-Tech University, Hangzhou 310018, China}

\author{Yunbo Zhang}
\email{ybzhang@zstu.edu.cn}
\affiliation{Department of Physics and Key Laboratory of Optical Field Manipulation of Zhejiang Province, Zhejiang Sci-Tech University, Hangzhou 310018, China}

\date{\today}

\begin{abstract}
The collective excitation modes in quantum droplets trapped in a two-dimensional harmonic potential in the context of symmetric weakly interacting binary bosonic mixtures are studied. By utilizing the linearization technique, the time-dependent extended Gross-Pitaevskii equation, and a sum-rule approach with a variational approximation, the ground state properties and collective excitations of such a two-dimensional quantum system are investigated for various system parameters. We present comprehensive analysis and calculations on the effect of the confinement strength and anisotropy of the trapping potential, the number of atoms in the droplet, and the collective excitation modes. The radius of the droplet, as well as the chemical potential, is non-monotonically related to the number of atoms in the droplet, and the confinement tends to shift the minimum values towards the ideal gas limit. The excitation frequency peaks, which are prominent in a self-bounded droplet, become less pronounced and smoother when subjected to a strong trapping potential. The sum-rule approach fails to reproduce the breathing mode frequency for a moderate number of atoms in a weak trapping potential, however, works perfectly well in a strong confinement. It was found that the anisotropy in the trap eliminates the degeneracy between the quadrupole and scissors modes that occurs in an isotropic trap, causing the frequencies of these two modes to immediately diverge from each other for any degree of anisotropy. These findings provide valuable insights into the unique characteristics and behavior of quantum droplets, offering potential implications for future research and applications in the dynamic behaviors of intriguing quantum droplets.
\end{abstract}

\maketitle

%%%%%%%%%%%%%%%%%%%%%%%%%%%%%%%%%%%%%%%%%%%%%%%%%%%%%%
%%%%%%%%%%%%%%%%%%%%%%%%%%%%%%%%%%%%%%%%%%%%%%%%%%%%%%
\section{INTRODUCTION}

The quantum droplet stands as an intriguing quantum state that has drawn significant attention within the field of ultracold atoms over the past decade~\cite{kartashov2019frontiers,bottcher2020new,luo2021new}. Initially conceptualized in the context of weakly interacting binary bosonic mixtures, Petrov extended the system to the beyond-mean-field (BMF) level in free space.~\cite{petrov2015droplets,petrov2016lowdimensional,petrov2023beyond}. Remarkably, he found that this nontrivial quantum droplet state can achieve stability through the influential repulsive Lee-Huang-Yang (LHY) correction energy~\cite{LHY1957}, counteracting the tendency to collapse under the attractive mean-field interaction alone. For instance, in three-dimensional bosonic mixtures, the attractive mean-field energy density is proportional to the square of density as $\propto n^2$ which can be properly compensated by the repulsive LHY energy $\propto n^{5/2}$, reaching an elegant balance to stabilize the quantum droplets. Meanwhile, this self-bound quantum droplet state is also predicted to exist in quantum dipolar gases in the presence of the BMF quantum fluctuations~\cite{baillie2016self,wachtler2016quantum,bisset2016ground,saito2016path}. Nowadays, this intriguing quantum state has been successfully achieved in both binary~\cite{cabrera2018quantum,semeghini2018self,cheiney2018bright,dErrico2019observation,ferioli2019collisions,skov2021observation} and dipolar~\cite{kadau2016observing,ferrier2016observation,chomaz2016quantum,schmitt2016self,tanzi2019observation,bottcher2019dilute} Bose gases. These experimental realizations have provided great platforms to investigate the underlying physics behind this fascinating phenomenon. They have also stimulated extensive and intensive theoretical studies of quantum droplets in bosonic mixtures, such as soliton-droplet~\cite{cappellaro2018collective} or liquid-gas~\cite{he2023quantum} transitions, vortex droplets~\cite{kartashov2018three,li2018two,lee2018excitations,cidrim2018vortices,kartashov2019metastability,otajonov2020variational},
effects of thermal fluctuations~\cite{guebli2021quantum} and dimensional crossover~\cite{zin2018quantum}, collisional and rotational properties~\cite{cikojevic2021dynamics,mithun2021statistical,hu2022collisional,yang2023two,tengstrand2022droplet,nikolaou2023rotating}, and collective excitations~\cite{astrakharchik2018dynamics,otajonov2019stationary,sturmer2021breathing,tylutki2020collective,hu2020collective,dong2022internal,nie2023dynamically}, etc.

In weakly interacting bosonic mixtures within the BMF level, the Bose-Einstein condensate (BEC) in free space will experience from a typical Gaussian-like state to a self-bound quantum droplet state possessing a flat-top-structure density in one, two, and three dimensions~\cite{petrov2015droplets,petrov2016lowdimensional,astrakharchik2018dynamics,li2018two,hu2020collective}. Several measurable observables like the radius of droplet, chemical potential, and collective oscillations will exhibit characteristic non-monotonic behavior with respect to the number of atoms. For instance, the mean-square size or breathing mode frequency in this mixture will exhibit typical behaviors of varying gradually for small sized droplet, significantly changing to a minimum or maximum at intermediate number of atoms, and then turning back at relatively large droplet~\cite{jorgensen2018dilute,astrakharchik2018dynamics,hu2020collective,tylutki2020collective,sturmer2021breathing}. These nontrivial properties can be well characterized by employing various approaches such as the widely used extended Gross-Pitaevskii equation (GPE) with BMF energy term, the quantum Monte Carlo method~\cite{parisi2019liquid,parisi2020quantum}, the Gaussian state theory~\cite{wang2020theory,pan2022quantum}, and a bosonic pairing theory~\cite{hu2020consistent}.

In current ultracold atomic experiments of quantum droplets, the external harmonic trapping potential is an necessary tool employed to induce this exotic state~\cite{cabrera2018quantum,semeghini2018self,cheiney2018bright,dErrico2019observation,ferioli2019collisions,skov2021observation}. It would be interesting to explore the interplay between the mean-field and LHY interactions alongside this trapping potential. Theoretically, Cappellaro {\it et al.} considered this one-dimensional binary Bose mixtures in harmonic confinements using a variational approach with a Gaussian {\it Ansatz}, and found a sharp difference in the frequencies of collective modes in the soliton and droplet regimes~\cite{cappellaro2018collective}. Later, Hu and Liu conducted a comprehensive investigation of three-dimensional quantum droplets in a spherical harmonic trap and found the droplet-to-gas transition as well as the excitation spectrum are significantly affected by the harmonic trap~\cite{hu2020collective}. The impact beyond LHY physics, especially due to intercomponent correlations, is extended to one-dimensional heteronuclear quantum droplet in a harmoic trap and the static density distributions as well as quench dynamics are studied~\cite{mistakidis2021formation}. By means of the construction of an exact analytical model from the one-dimensional extended GPE, the structure and dynamics of trapped quantum droplets and the interesting droplet-soliton phase transition has been investigated~\cite{pathak2022dynamics}. Recently, interesting nonlinear dynamic behaviors are examined for the excited states in a quasi-one-dimensional system with a parabolic trap~\cite{Zezyulin2023quasi} and the excitation frequencies of the breathing mode in one-dimensional trapped quantum droplets are calculated numerically by a linearization technique~\cite{du2023ground}. Nonetheless, the investigation of collective exicitations of quantum droplets in two-dimensional external traps is still elusive, and the effect of anisotropy in the harmonic trap is not clear.

This paper aims to fill this gap and we investigate the effect of harmonic traps on the ground state properties and collective excitation spectrum of the quantum droplet in a two-dimensional trap in the context of weakly interacting binary Bose gas, utilizing three different methods, both numerical and analytical. The ground state wavefunction is obtained through the extended GPE, and the density profile, droplet width, and chemical potential are thoroughly examined at various strengths of the trapping potential and number of atoms. We then linearize the time-dependent extended GPE and derive the coupled equations for small fluctuations, with which the frequency of low-lying collective excitation modes as a function of normalized number of atoms can be numerically calculated. To further verify our numerical results for the excitation frequency, we generalize the time-dependent extended GPE with designed perturbations to excite specific collective modes such as dipole, monopole (breathing), quadrupole, and scissors modes. In addition, a sum-rule approach with a Gaussian and super-Gaussian variational {\it Ansatz} is developed to obtain the analytic expression for the frequency of breathing mode. In the end, by tuning the anisotropy in the harmonic trap for different strengths of trapping potential, the anisotropic impact on the density distribution as well as the excitation frequency are explored.

The rest of the paper is organized as follows. In Sec.~\ref{sec:theo}, we start with the model and derive the extended GPE for this two-dimensional symmetric bosonic mixture. Three primary methods, i.e., the linearization technique, time-dependent extended GPE with perturbations, and a sum-rule approach based on variational approximation will be briefly introduced. In Sec.~\ref{sec:results}, we apply these methods to calculate the ground-state density distribution, width of the quantum droplet, chemical potential, and excitation frequency of low-lying collective modes in the presence of harmonic trap. The effects of the number of atoms, confinement strength, and anisotropy of the harmonic traps on quantum droplets are then comprehensively discussed. Conclusions and outlooks are given in Sec.~\ref{sec:conclusions}.

%%%%%%%%%%%%%%%%%%%%%%%%%%%%%%%%%%%%%%%%%%%%%%%%%%%%%%
%%%%%%%%%%%%%%%%%%%%%%%%%%%%%%%%%%%%%%%%%%%%%%%%%%%%%%
\section{THEORETICAL FRAMEWORK} \label{sec:theo}
In this section, we will provide a brief overview of three primary methods utilized for qualitatively and quantitatively studying the collective excitations of quantum droplets in this work. These methods are based on the extended Gross-Pitaevskii equation (GPE), which can well describe the static and dynamical properties of quantum droplets in binary bosonic mixtures.

%%%%%%%%%%%%%%%%%%%%%%%%%%%%%%%%%%%%%%%%%%%%%%%%%%%%%%
\subsection{Model Hamiltonian and the extended Gross-Pitaevskii equation}
We start from a homogeneous two-dimensional weakly interacting Bose-Bose mixture with the same atomic mass $m_1=m_2=m$, and the state of quantum droplets can be stabilized by the beyond-mean-field correction in the Bogoliubov approximation. The corresponding energy density functional is written as~\cite{petrov2016lowdimensional}
\begin{equation}
    \varepsilon(\psi^*,\psi) = \frac{\hbar^2|\nabla\psi|^2}{2m} + \frac{8\pi\hbar^2|\psi|^4}{m \ln^2{(a_{12}/a}) } \ln \frac{|\psi|^2}{en_\mathrm{eq}},
\end{equation}
with the two-dimensional intra- and inter-species scattering length $a_{11}=a_{22}=a$ and $a_{12}$, and the equilibrium density~\cite{petrov2016lowdimensional}
\begin{equation}
    n_\mathrm{eq}=\frac{e^{-2\gamma-\frac{3}{2}}}{2\pi}\frac{\ln(a_{12}/a)}{aa_{12}},
\end{equation}
is determined under the zero-pressure condition with the Euler constant $\gamma \approx 0.5772$.

For trapped droplets, we therefore turn on the external trapping potential $V_\mathrm{ext}({\bf r})$ with two-dimensional coordinate ${\bf r}\equiv(x,y)$, minimize the total energy functional with respect to the Bose field, and derive the so-called extended GPE for such a symmetric bosonic mixture in harmonic traps, i.e.,~\cite{petrov2015droplets,petrov2016lowdimensional,hu2022collisional}
\begin{equation} \label{eGPE-2D}
    i\hbar\frac{\partial}{\partial t} \psi({\bf r},t) = \left[H_\mathrm{ho} + \frac{8 \pi\hbar^2|\psi|^2}{m\ln^2 (a_{12}/a)}\ln \frac{|\psi|^2}{\sqrt{e}n_\mathrm{eq}} \right]\psi({\bf r},t),
\end{equation}
with $H_\mathrm{ho}=-\frac{\hbar^2\nabla^2}{2m}+V_\mathrm{ext}({\bf r})$ being the single-particle Hamiltonian of a harmonic oscillator. The two-dimensional wave function of the condensate $\psi(x,y)$ is applicable for both spin components and satisfies the normalization condition $\iint|\psi(x,y)|^2 dxdy=N$. In this work, the external harmonic trapping potential is adopted as $V_\mathrm{ext}(x,y)=\frac{1}{2}m\omega_x^2(x^2+\kappa^2y^2)$ with the $x,y$-axis frequencies $\omega_{x,y}$ and an anisotropy parameter $\kappa\equiv\omega_y/\omega_x$ is introduced to describe the anisotropy of the harmonic trap.

Following the conventional procedure, we introduce a set of characteristic units and rewrite the extended GPE into a dimensionless form. The units adopted in this work are given by
\begin{eqnarray}
    {\psi_0} &\equiv& \sqrt{\sqrt{e}n_\mathrm{eq}}=\sqrt{\frac{e^{-2\gamma-1}\ln(a_{12}/a) }{2\pi a_{12}a}}, \nonumber \\ 
    {r_0}&=& \sqrt{\frac{\ln(a_{12}/a)a_{12}a}{4e^{-2\gamma-1}}}, \nonumber \\
    {E_0} &=&\frac{4e^{-2\gamma-1}\hbar^2}{m\ln(a_{12}/a)a_{12}a}, \\
    {t_0} &=& \frac{m\ln(a_{12}/a)a_{12}a}{4\hbar e^{-2\gamma-1}}, \nonumber \\
    {N_0} &=& \psi_0^2r_0^2={\ln^2(a_{12}/a)}/{8\pi}. \nonumber
\end{eqnarray}
In terms of these units, i.e., $\tilde{t}=t/{t_0}, \tilde{x}=x/{r_0},\tilde{y}=y/{r_0},\tilde{\psi}=\psi/{\psi_0},\tilde{N}=N/{N_0}$, the dimensionless form of the extended GPE can be simplified as 
\begin{equation} \label{EGPe_2D_dimensionless}
    i\partial_{t} \psi=\left( H_\mathrm{ho}+ |\psi|^2 \ln |\psi|^2\right) \psi,
\end{equation}
%\begin{equation} \label{EGPe_2D_dimensionless}
%    i\partial_{\tilde{t}} \tilde{\psi}=\left( \tilde{H}_\mathrm{ho}+ |\tilde{\psi}|^2 \ln |\tilde{\psi}|^2\right) \tilde{\psi},
%\end{equation}
%with $\tilde{H}_\mathrm{ho}=-{\nabla_{\tilde{x}}^2}/{2} +\lambda_{x}^2(\tilde{x}^2+\kappa^2\tilde{y}^2)/2$ and $\lambda_{x}=\hbar\omega_x/E_0$ 
where we have omitted the tilde symbols hereafter for simplicity. Here $H_\mathrm{ho}=-{\nabla_{x}^2}/{2} +\lambda_{x}^2(x^2+\kappa^2y^2)/2$ and the dimensionless coefficient $\lambda_{x}=\hbar\omega_x/E_0$ describes the ratio of the strength of the trapping potential over the characteristic energy unit which will be utilized in the later section. Note that the normalization condition is exactly the same before and after the dimensionless procedure.

%%%%%%%%%%%%%%%%%%%%%%%%%%%%%%%%%%%%%%%%%%%%%%%%%%%%%%
\subsection{Linearization technique}

In order to investigate the collective excitations in a weakly interacting Bose gas, the celebrated Bogoliubov approximation or the linearization technique is usually utilized~\cite{nie2023dynamically,hu2020collective,tylutki2020collective,dong2022internal,du2023ground}. In practice, one may linearize the time-dependent Bose field to be a superposition of a static classic ground-state field and small-amplitude oscillations around the ground state. The properties of collective excitations can be well captured by further studying the oscillations in the condensate. Therefore, by employing this technique, the Bose field in the extended GPE Eq.~\eqref{EGPe_2D_dimensionless} can be linearized by
\begin{equation} \label{eq:linearized}
    \psi({\bf r},t)=e^{-i\mu_g t }\left\{ \psi_g({\bf r})+\sum_j \left[ u_j({\bf r}) e^{-i\omega_j t} +v_j^*({\bf r})e^{i\omega_j t} \right]  \right\},
\end{equation}
with the static ground-state wave function $\psi_g$ and the amplitude wave functions $u_j, v_j$ of the $j$-th mode oscillations. Here, $\omega_j$ is the wanted excitation energy or frequency of the $j$-th oscillating mode. Hence, after substituting the linearized expression of the time-dependent field $\psi({\bf r},t)$ back to Eq.~\eqref{EGPe_2D_dimensionless} and keeping up to the first-order terms in the small fluctuations $u_j$ and $v_j$, we obtain the static equation for the condensate wave function
\begin{equation} \label{eq:gp-static}
    \mu_g \psi_g({\bf r}) = \left[H_\mathrm{ho} + n_g\ln{n_g} \right]\psi_g({\bf r}),
\end{equation}
with the chemical potential $\mu_g$ and the density $n_g=|\psi_g|^2$ of the condensate, as well as the coupled equations for the small fluctuations
\begin{equation} \label{BdG}
    \begin{pmatrix}
        \mathcal{L} &\mathcal{M} \\ -\mathcal{M}^* &-\mathcal{L}
    \end{pmatrix}\begin{pmatrix}
        u_j \\ v_j
    \end{pmatrix} = \omega_j\begin{pmatrix}
        u_j \\ v_j
    \end{pmatrix},
\end{equation}
where we have introduced
\begin{subequations}
\begin{eqnarray}
	\mathcal{L} &=& H_\mathrm{ho}-\mu_g+2 n_g\ln n_g +  n_g,\\
	\mathcal{M} &=& n_g\ln n_g + n_g.
 \end{eqnarray}
\end{subequations}

In the following we first solve the modified static GPE Eq.~\eqref{eq:gp-static} and obtain the ground-state wave function $\psi_g(\bf r)$ as well as the associated static properties at chosen parameters. After this, the linearized coupled equations Eq.~\eqref{BdG} are further solved, and the crucial excitation spectra or frequencies $\omega_j$ of specific oscillating modes are then calculated as functions of various parameters to demonstrate the dynamic properties of the droplet. In our numerical calculations, we have adopted a sine-pseudospectral method~\cite{bao2006efficient,gao2020numerical} within an imaginary-time propagation method to solve the equations above.

%%%%%%%%%%%%%%%%%%%%%%%%%%%%%%%%%%%%%%%%%%%%%%%%%%%%%%
\subsection{Time-dependent extended Gross-Pitaevskii equation with perturbations}

In addition to the linearization technique, an alternative way to introduce small perturbations into the static system is to excite specific collective modes such as the dipole, monopole, and quadrupole modes, which are of great interest and feasibly accessible in experiments~\cite{cheiney2018bright,cabrera2018quantum,dErrico2019observation,schmitt2016self,ferrier2016observation,chomaz2016quantum,tanzi2019observation,bottcher2019dilute}. In this work, we focus on four particular collective modes in the two-dimensional geometry, i.e., the dipole, monopole (or breathing), quadrupole, and scissors modes, and choose different types of time-dependent external traps $V_\mathrm{ext}({\bf r},t)$ to excite the corresponding oscillation modes. 

We consider the following time-dependent external trapping potential
\begin{equation} \label{breathing}
	V_\mathrm{ext}({\bf r},t)=\left\{
	\begin{array}{ccc}
		&V_0({\bf r}), & {t \leq 0},\\
		&V_1({\bf r}), & {t > 0},
	\end{array} \right.
\end{equation}
with the above static harmonic trapping potential $V_0({\bf r})=\lambda_x^2(x^2+\kappa^2y^2)/2$, and a slightly modified one $V_1({\bf r})$. We first obtain the ground-state wave function under the potential $V_0$ at $t=0$ and suddenly change to $V_1$. By employing the time-dependent extended GPE Eq.~\eqref{eGPE-2D}, we calculate the evolution of the wave function and construct the time-dependent observables for any later time $t>0$. Furthermore, the time dependence of these observables will be carefully analyzed by means of the Fourier transformation, and the oscillation frequencies of the collective modes are successfully extracted.

In practice, the dipole, breathing, quadrupole, and scissors modes can be excited by employing the following modified trapping potential~\cite{ueda2010fundamentals,pitaevskii2016bose}
\begin{widetext}
    \begin{equation} \label{eq:V1}
	V_{1}({\bf r}) = \left\{
		\begin{array}{rcl}
		&\frac{1}{2}\lambda_x^2 \left[ (x-\delta x)^2+\kappa^2y^2\right], &\text{for dipole mode (d.m.),} \\
		&\frac{1}{2}(\chi\lambda_x)^2(x^2+\kappa^2y^2), &\text{for breathing mode (b.m.),} \\
		&\frac{1}{2}\lambda_x^2\left[(\chi x)^2+\kappa^2y^2\right], &\text{for quadrupole mode (q.m.),} \\			  
            &\frac{1}{2}\lambda_x^2\bigl[(x\cos\theta-y\sin\theta )^2+\kappa^2(x\sin\theta +y \cos \theta)^2\bigr], &\text{for scissors mode (s.m.).} \\
		\end{array} \right.
    \end{equation}
\end{widetext} 
Here, $\delta x$, $\chi$, and $\theta$ are small variables relative to zero or unity for introducing perturbations in the original trapping potentials. Explicitly, we move the potential along the $x$ or $y$ axis, i.e., $x \rightarrow  x-\delta x$ and $y \rightarrow  y-\delta y$ in Eq.~\eqref{eq:V1}, and the condensate will commit the $x$- or $y$-axis dipole oscillations around the origin under the modified potential. This center-of-mass motion can be characterized by calculating the time-dependent mean radius $\langle x\rangle(t)$ or $\langle y\rangle(t)$. The oscillation frequency of the dipole mode can be straightforwardly obtained by applying the Fourier transformation. According to Kohn's theorem, the dipole-mode frequency is the same as that of the trapping potential, i.e., $\omega_\mathrm{d.m.}=\omega_{x,y}$, independent of the interactions in a weakly interacting Bose gas~\cite{Kohn1961cyclotron}. For the monopole-mode oscillation, one can either tune isotropically the trapping frequency [i.e., $\lambda_x \rightarrow \chi\lambda_x$ in Eq.~\eqref{eq:V1}], or change slightly the normalization (i.e., $N \rightarrow \chi N$ as in Ref.~\cite{astrakharchik2018dynamics,sturmer2021breathing}), leading the atomic cloud to exhibit a breathing-like oscillation with the system expanding and contracting isotropically. Similarly, we can calculate the time-dependent mean squared radius $\langle r^2\rangle(t)$ to depict this monopole or breathing mode, and the corresponding oscillation frequency is then obtained by the Fourier analysis. Besides, we change merely the $x$-axis trapping frequency via $x \rightarrow \chi x$ in Eq.~\eqref{eq:V1} to excite the quadrupole mode, and add a small twist angle $\theta$ via 
\begin{equation}
    \begin{pmatrix} x'\\y' \end{pmatrix}
    = \begin{pmatrix}
        \cos\theta &-\sin\theta \\
        \sin\theta &\cos\theta
    \end{pmatrix}\begin{pmatrix} x\\y \end{pmatrix}
\end{equation}
to excite the scissors mode~\cite{guery1999scissors,marago2000observation}.
These two perturbations described above, on the other hand, usually give rise to the deformation of the atomic cloud and will excite simultaneously both the quadrupole and scissors modes. In general, one can use the time-dependent averaged observables $\frac{\langle x^2-y^2 \rangle}{\langle x^2+y^2 \rangle}(t)$ or $\langle xy\rangle(t)$ to characterize the evolution of deformation in the atomic cloud. After applying the Fourier transform, there are usually two obvious peaks in the amplitude-frequency plot associated to these two coupled oscillation modes. To further distinguish these two modes, we will carefully inspect the time evolution of the density profile and identify the pattern in the oscillations.
\begin{figure*}[ht]
\centering
\includegraphics[width=0.96\textwidth]{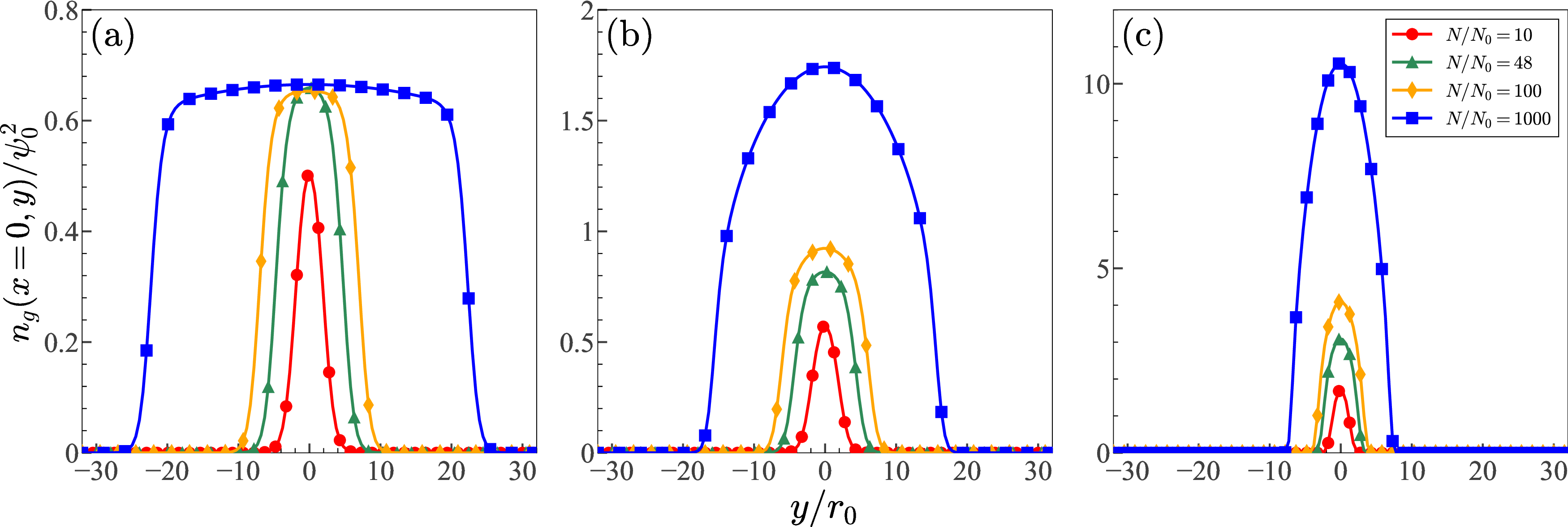}
\caption{Typical density profiles $n_g(0,y)/\psi_0^2$ of the droplet at different normalization $N/N_0$ for three values of the trapping strength $\lambda_x\equiv\hbar\omega_x/E_0$ with (a) $\lambda_x=0.01$, (b) $\lambda_x=0.1$, and (c) $\lambda_x=1$ in an isotropic two dimensional trap $\kappa=1$. Here, the colorfedlines with dots, triangles, diamonds, and squares correspond to $N/N_0=$10, 48, 100, and 1000, respectively.}
\label{fig1}
\end{figure*}

%%%%%%%%%%%%%%%%%%%%%%%%%%%%%%%%%%%%%%%%%%%%%%%%%%%%%%
\subsection{Sum-rule approach}

Following Ref.~\cite{hu2020collective}, we may also introduce a sum-rule approach to get a preliminary qualitative description of the breathing mode of the quantum droplets in symmetric Bose mixtures. In previous studies~\cite{astrakharchik2018dynamics,hu2020collective,hu2022collisional,petrov2016lowdimensional,sturmer2021breathing,du2023ground}, the ground-state wave function of quantum droplets may differ significantly from a conventional Gaussian function and the density distribution may exhibit a flat-top shape. Therefore, in addition to a conventional Gaussian variational {\it Ansatz}, we adopt a two-dimensional super-Gaussian (sG) variational {\it Ansatz}~\cite{karlsson1992optical,sturmer2021breathing,tsoy2006dynamical,otajonov2019stationary,baizakov2011variational}
\begin{equation}
    \psi^\mathrm{sG}(r)=A \exp\left[-\frac{1}{2}\left(\frac{r}{R}\right)^{2l}\right],\label{sG}
\end{equation}
with the normalized coefficient $A=\sqrt{\frac{Nl}{\pi R^2 \Gamma(\frac{1}{l})}}$ and the radius $r\equiv\sqrt{x^2+y^2}$. Here, $R$ and $l$ are the variational width and power, respectively. It's easy to see that the Gaussian {\it Ansatz} is recovered at $l=1$. These variational parameters can be then determined by minimizing the dimensionless total energy
\begin{equation}
E_\mathrm{tot}=\int d{\bf r} \left[\frac{(\nabla \psi)^2}{2}+\frac{1}{2}\lambda_x^2r^2\psi^2+\frac{1}{2}|\psi|^4\ln\frac{|\psi|^2}{\sqrt{e}} \right].
\end{equation}
After substituting the variational {\it Ansatz} (\ref{sG}) into the total energy and performing some straightforward algebras, we obtain the energy per particle
\begin{eqnarray}
\epsilon_\mathrm{tot} &\equiv& \frac{E_\mathrm{tot}}{N}=\frac{l^2}{2R^2\Gamma(\frac{1}{l})}+\frac{\lambda_x^2R^2}{2}\frac{\Gamma(\frac{2}{l})}{\Gamma(\frac{1}{l})}+ \nonumber
\\
&&\frac{Nl}{2^{\frac{1}{l}+1}\pi R^2 \Gamma(\frac{1}{l})}
\left[\ln(\frac{Nl}{\sqrt{e}\pi R^2 \Gamma(\frac{1}{l})})-\frac{1}{2l}  \right].
\end{eqnarray}
After minimizing this energy density with respect to the variational parameters, we get a relation
\begin{equation} \label{lambda_x}
   \lambda_x^2 = \frac{1}{R^4}\left[\frac{l^2}{\Gamma(\frac{2}{l})} +  \frac{Nl}{2^{\frac{1}{l}}\pi \Gamma(\frac{2}{l})}
\left(\ln\frac{\sqrt{e}Nl}{\pi R^2 \Gamma(\frac{1}{l})}-\frac{1}{2l}  \right)  \right], 
\end{equation}
with which we can numerically determine the width $R$ and power $l$ in the droplet wave function for a given set of parameters $(\lambda_x,N)$. In particular, we can use this relation to further calculate the associated breathing-mode frequency $\omega_\mathrm{b.m.}$ by employing the sum-rule approach~\cite{stringari1996collective,dalfovo1999theory,menotti2002collective,hu2020collective}
\begin{equation}
\omega_\mathrm{b.m.}^2=-2\frac{\left\langle r^2 \right\rangle }{\partial\left\langle r^2 \right\rangle /\partial \lambda_x^2}=-R\frac{\partial \lambda_x^2 }{\partial R}.
\end{equation}
After some simple algebras, we get an analytic expression for the frequency of breathing mode as
\begin{equation} \label{omega_b_sG}
    \omega_\mathrm{b.m.,sG}^2=4\left(\lambda_x^2+\frac{Nl}{2^{\frac{1}{l}+1}\pi R^4 \Gamma(\frac{2}{l})}\right),
\end{equation}
in terms of $R$ and $l$ which can be numerically calculated. Similarly, by setting $N=0$ in Eq.~\eqref{lambda_x} and Eq.~\eqref{omega_b_sG}, we may recover the width $R=\lambda_x^{-1/2}\cdot r_0=\sqrt{\frac{\hbar}{m\omega_x}}$ and the breathing-mode frequency $\omega_\mathrm{b.m.}=2\lambda_x\cdot E_0/\hbar=2\omega_x$ for a non-interacting gas~\cite{pitaevskii2016bose}. Particularly at $l=1$, we obtain the breathing-mode frequency of a two-dimensional weakly interacting Bose mixture starting from a Gaussian (G) {\it Ansatz}, written as
\begin{equation} \label{omega_b_G}
\omega_\mathrm{b.m.,G}^2=4\left(\lambda_x^2+\frac{N}{4\pi R^4 }\right)=\frac{4}{R^4}\left( 1+\frac{N}{2\pi}\ln \frac{\sqrt{e}N}{\pi R^2}\right).
\end{equation}
It is worth noting that, we have also tried the variational approximation together with the Euler-Lagrange equations as in Refs.~\cite{astrakharchik2018dynamics,sturmer2021breathing} to calculate the breathing-mode frequency for such a two-dimensional harmonically trapped bose mixture, and the results are the same as in Eq.~\eqref{omega_b_sG} and Eq.~\eqref{omega_b_G}.

%%%%%%%%%%%%%%%%%%%%%%%%%%%%%%%%%%%%%%%%%%%%%%%%%%%%%%
%%%%%%%%%%%%%%%%%%%%%%%%%%%%%%%%%%%%%%%%%%%%%%%%%%%%%%
\section{RESULTS AND DISCUSSIONS} \label{sec:results}

In this section, we first present the stationary properties of the condensate as well as the excitation spectrum with respect to the normalization $N$ for such a two-dimensional bose mixture in an isotropic harmonic trap, and then discuss the effect of anisotropy in harmonic traps on the frequencies of various collective modes by tuning the anisotropy parameter $\kappa$. In our numerical calculations, we choose a two-dimensional box geometry with a length $L_{xy}=64$ in each axis and a spatial interval $\Delta h_{xy}=0.5$.

\begin{figure*}[ht]
\includegraphics[width=0.32\textwidth]{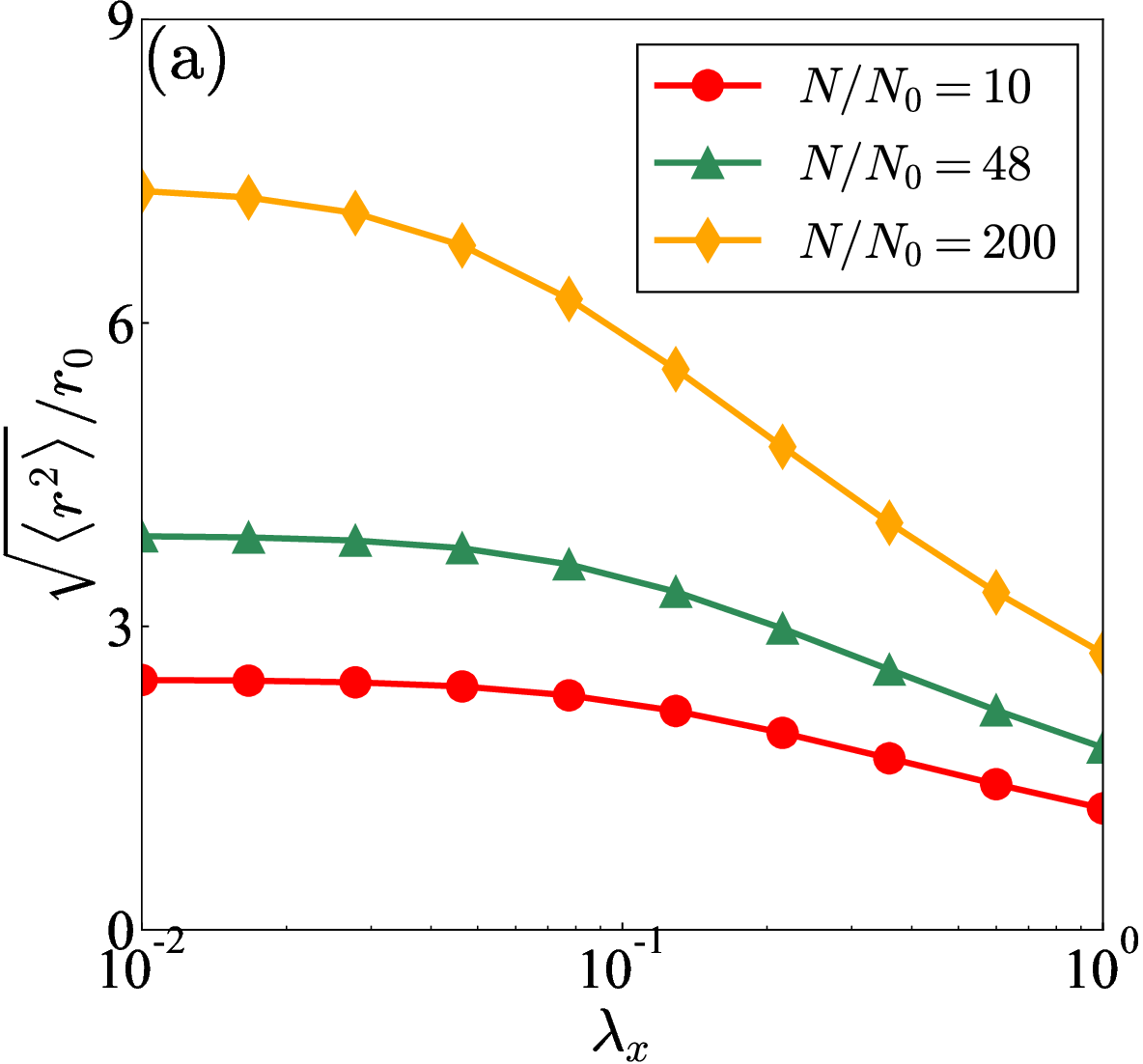}
\includegraphics[width=0.32\textwidth]{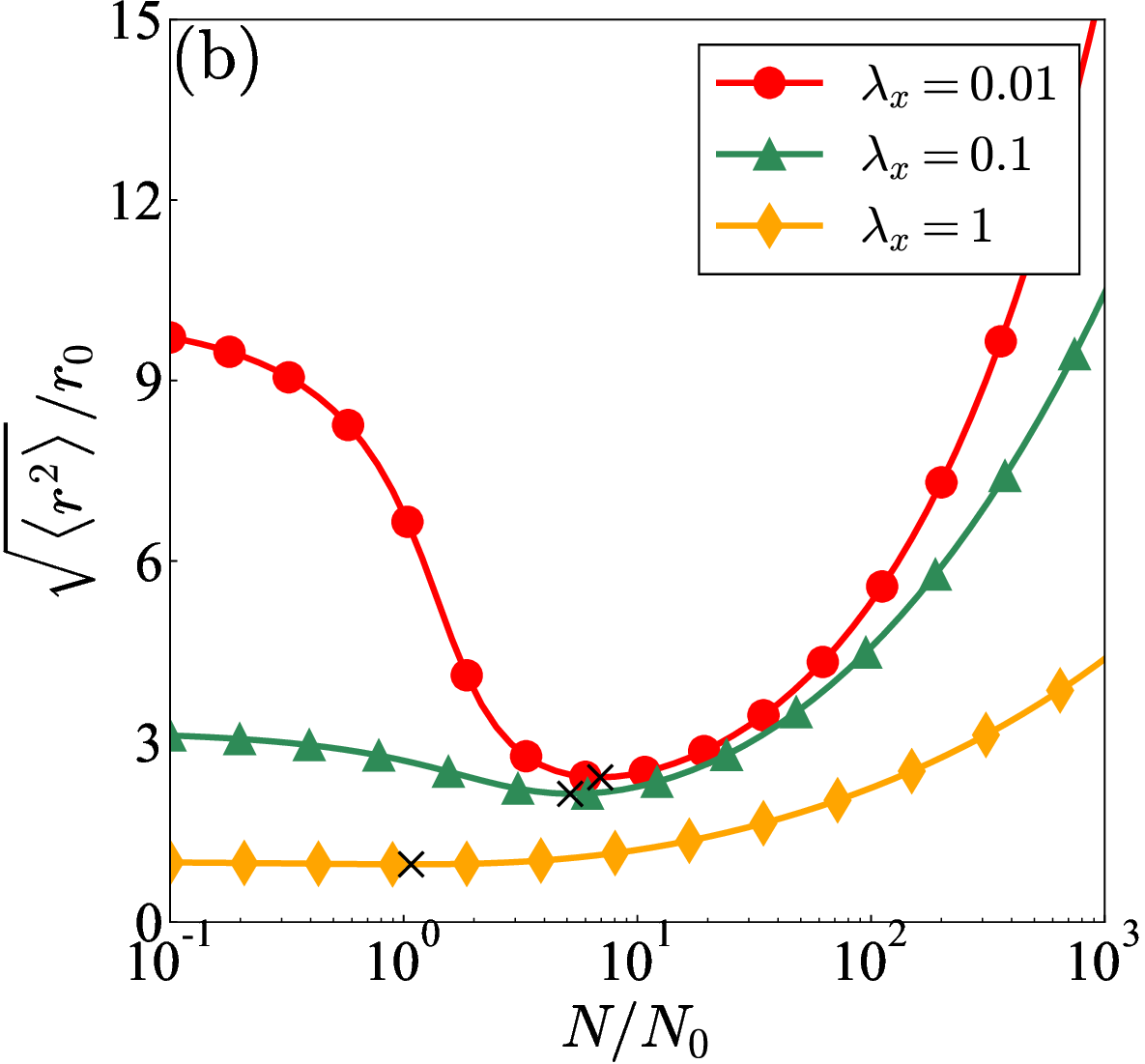}
\includegraphics[width=0.32\textwidth]{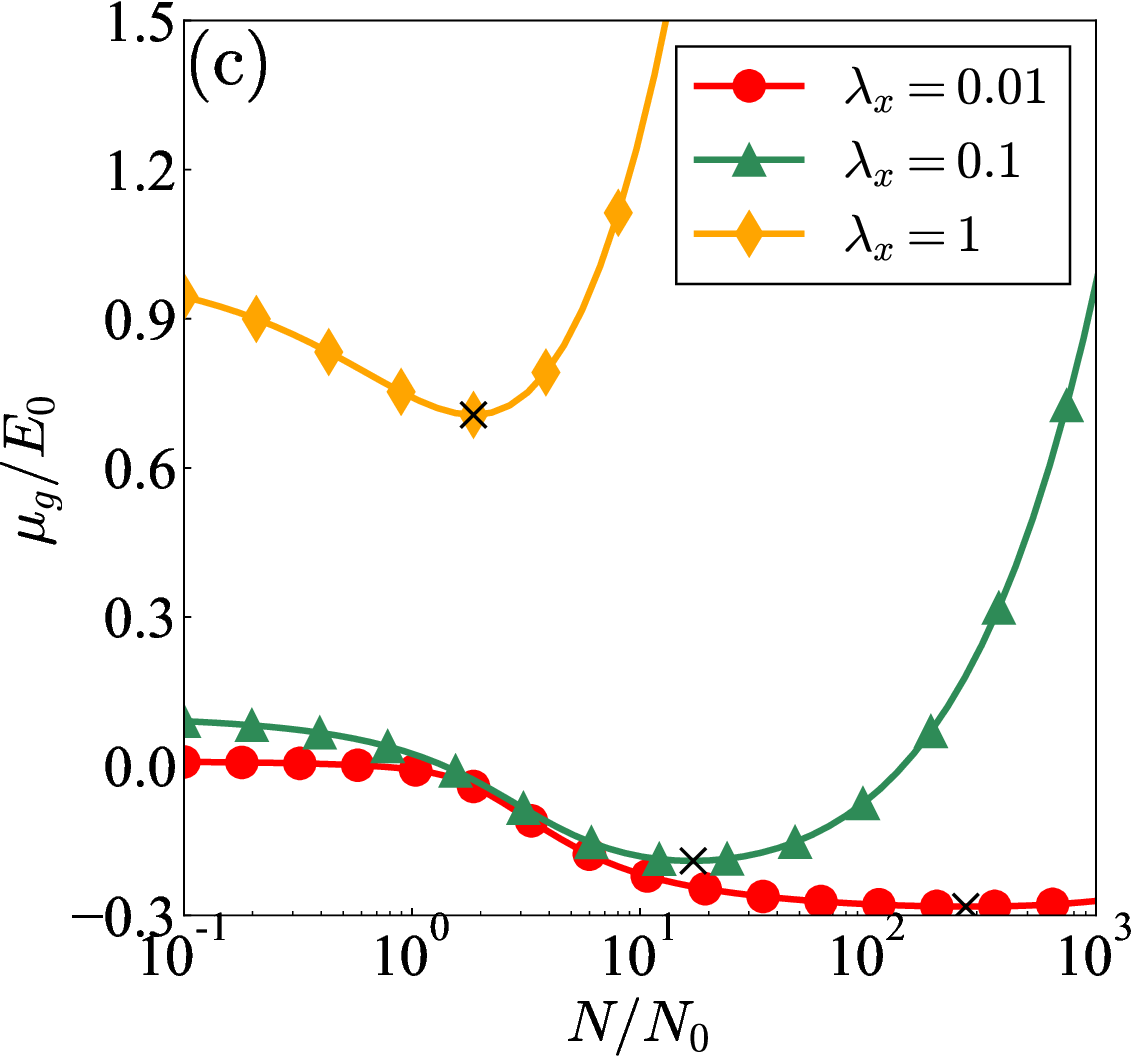}
\caption{(a) Root-mean-square size $\sqrt{\langle r^2\rangle}/r_0$ of the droplet as a function of the trapping strength $\lambda_x\equiv \hbar\omega_x/E_0$ at three typical values of $N/N_0$. (b, c) Root-mean-square size $\sqrt{\langle r^2\rangle}/r_0$ and chemical potential $\mu_g/E_0$ of the droplet, as functions of the normalization $N/N_0$ at three typical values of $\lambda_x$. The crosses label the minimum of each curve. The other parameters are the same as in Fig.~\ref{fig1}.}
\label{fig2}
\end{figure*}

%%%%%%%%%%%%%%%%%%%%%%%%%%%%%%%%%%%%%%%%%%%%%%%%%%%%%%
\subsection{Stationary properties}

We first discuss the stationary properties of this two-dimensional harmonically trapped droplet by numerically solving the static extended GPE in Eq.~\eqref{eq:gp-static}. In Fig.~\ref{fig1}, the typical density distributions $n_g(x=0,y)$ at different normalized number of atoms $N$ are presented for three typical values of the trapping strength $\lambda_x$. The subplots from left to right denote three trapping strengths $\lambda_x=0.01$, 0.1, and 1, while the red, green, orange, and blue lines with various symbols denote four values of normalized number of atoms $N/N_0=10$, 48, 100, and 1000, respectively. In Fig.~\ref{fig1} (a) with a negligible trapping potential, we reproduce the results of quantum droplets as in previous works for one~\cite{petrov2016lowdimensional,astrakharchik2018dynamics,tylutki2020collective,parisi2020quantum,du2023ground}, two~\cite{li2018two,otajonov2020variational,sturmer2021breathing}, and three dimensions~\cite{petrov2015droplets,hu2020collective,guebli2021quantum}, i.e., the density distribution evolves from a Gaussian-like shape to a flat-top structure with a saturated density as the normalized $N$ exceeds a critical value. In two dimensions, the critical value for the appearance of the self-trapped quantum droplets is about $N_c\approx48N_0$ as shown in green line in the figure~\cite{li2018two}. This intriguing stable self-bounded behavior of the quantum droplet state is a direct consequence of the interplay between attractive (repulsive) mean-field interatomic interactions and quantum fluctuations in such a quantum system~\cite{petrov2015droplets}. As the strength of the trapping potential is set to be non-negligible as shown in Fig.~\ref{fig1} (b), the density distribution experiences a change from a Gaussian-like shape of ideal gases to a Thomas-Fermi distribution of the conventional Bose-Einstein condensate where the peak values of the density increases with the number of atoms. In this case, the external trapping potential serves to stabilize the atoms from collapsing against the negative energy term which is a combination of the interatomic interaction term and LHY energy correction. For sufficiently strong trapping potential as shown in Fig.~\ref{fig1} (c), the density profiles exhibit similar Gaussian-to Thomas-Fermi transition behavior as the particle number increases except that the density width of the droplet is further squeezed.

We turn to discuss the mean-square radius or size $\sqrt{\langle r^2\rangle}$ and the chemical potential $\mu_g$ of the droplet as has done for one dimensional droplet in our previous  work~\cite{du2023ground}. In Fig.~\ref{fig2} (a), the mean-square size $\sqrt{\langle r^2\rangle}/r_0$ is presented as a function of the strength of the harmonic trapping potential $\lambda_x$ at three typical values of normalized number of atoms $N/N_0=10$, $48$, and $200$, respectively. For these three values of atomic numbers, the radius monotonically decreases as $\lambda_x$ increases since stronger trapping potential tends to confine the droplet more tightly and thus relatively reduce the size of the condensate. In the weak trapping potential limit (i.e., $\lambda_x\to0$, left axis), the radius of $N/N_0=200$ is significantly larger than for small atomic numbers and shrinks more drastically with the external confinement trap. This is because once the number of atoms exceeds the critical value (i.e., about $48N_0$), the quantum droplet manifests its profile with a saturated peak density, and thus the width is remarkably larger. In this limit, the calculated radius read $\sqrt{\langle r^2\rangle}/r_0=2.45$, $3.90$, $7.33$ for $N/N_0=10$, $48$, $200$, respectively, which exactly recover the results in the self-bound case~\cite{sturmer2021breathing}. In contrast, in the relatively strong trapping potential limit (i.e., $\lambda_x\to1$, right axis), the droplet experiences a change from a Gaussian-like profile to a conventional Thomas-Fermi-type density distribution where all the peak densities are not saturated. Thus, the width can be well squeezed by the external potential and gradually increase with the number of atoms.

In Fig.~\ref{fig2} (b) and \ref{fig2} (c), the radius $\sqrt{\langle r^2\rangle}$ and the chemical potential $\mu_g$ are also shown as a function of normalized number of atoms $N$ at three values of $\lambda_x$, respectively. For $\lambda_x=0.01$, we find a prominent non-monotonic behavior in $\sqrt{\langle r^2\rangle}$ and $\mu_g$ shown in the red lines with filled circles, similar to the uniform case in our previous study on one-dimensional droplet~\cite{du2023ground}. As $N$ increases, the radius decreases sharply, reaches a minimum, and then turns to rise back again, while the chemical potential decreases gradually to be negative and tends to be approximately saturated with the minimum almost invisible. In the presence of an external potential ($\lambda_x=0.1$ and $1$, denoted by the green lines with filled triangles and the yellow lines with filled diamonds, respectively), the variation of the radius becomes much smoother with $N$, and the position of the minimum shown by the crosses moves towards the ideal gas limit $N\to0$ at relatively stronger trapping potential (e.g. $\lambda_x=1$), giving rise to a nearly-monotonic curve. Meanwhile, the trend of the chemical potential is also significantly altered. The tendency of monotonic decreasing in the chemical potential is strongly suppressed and for trapped droplet, i.e., it is easily turn the chemical potential into positive and the minimum, labeled by the crosses, emerges more prominently and moves gradually towards the ideal gas limit as the confinement $\lambda_x$ becomes more and more stronger. 

%%%%%%%%%%%%%%%%%%%%%%%%%%%%%%%%%%%%%%%%%%%%%%%%%%%%%%
\begin{figure*}[ht]
\centering
\includegraphics[width=0.96\textwidth]{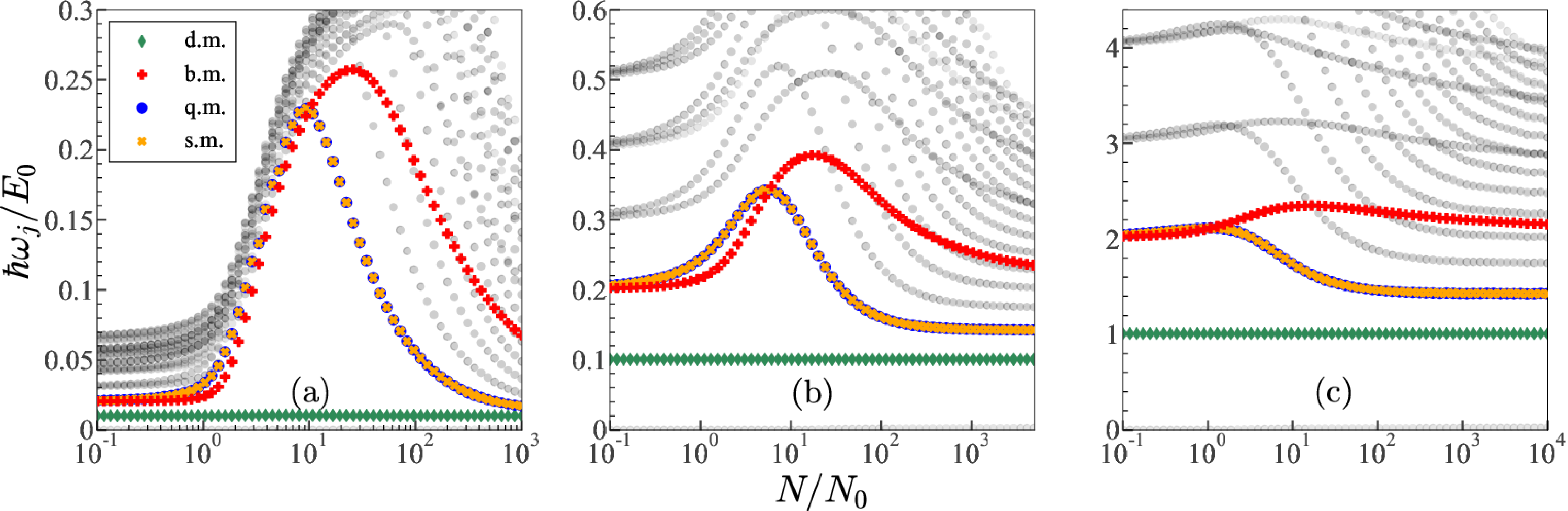}
\caption{Low-lying energy spectrum $\hbar\omega_j/E_0$ of the collective excitations as a function of the normalized number of atoms $N/N_0$ at three values of the trapping strength (a) $\lambda_x=0.01$, (b) $\lambda_x=0.1$, and (c) $\lambda_x=1$. Here, the green diamonds,  red pluses, blue dots, and orange crosses represent the frequencies of dipole mode (d.m.), breathing mode (b.m.), quadrupole mode (q.m.), and scissors mode (s.m.) respectively, while the light gray dots indicate other collective-mode frequencies. The other parameters are the same as in Fig.~\ref{fig1}.}
\label{fig3}
\end{figure*}

\subsection{Collective excitations}

In this section, we explore collective excitations in this two-dimensional trapped droplet and study the properties of the low lying collective modes, by utilizing diverse theoretical and numerical approaches introduced in Sec.~\ref{sec:theo}.

In Fig.~\ref{fig3}, the excitation frequencies $\omega_j$ of the low-lying collective modes are depicted as functions of the number of atoms $N$ at three values of trapping strength $\lambda_x=0.01$, $0.1$ and $1$ by numerically solving the coupled equations for fluctuations Eq.~\eqref{BdG}. The particle-emission threshold no longer exists due to the external trap~\cite{hu2020collective}. As a consequence, the related surface and bulk modes in the excitation spectra become bounded and have discrete mode frequencies denoted by the colored symbols indicated in the plots. In general, we find that all low-lying excitation frequencies except for the lowest dipole mode exhibit non-monotonic behavior, i.e., first gradually increasing to a maximum, then decreasing, and finally tending to saturate as $N$ increases. This intriguing behavior is consistent with the findings in the one- and three-dimensional cases with harmonic traps~\cite{du2023ground,hu2020collective}. In the limit $N\to0$ we recover the excitation frequencies $\hbar\omega_j=n\hbar\omega_x=n\lambda_x E_0$ ($n=1,2,3,\cdots$) of a quantum harmonic oscillator. In particular, four typical collective modes, i.e., the dipole, breathing, quadrupole, and scissors modes, discussed in tprevious section, are shown by green diamonds, red pluses, blue dots, and orange crosses, respectively. In practice, to distinguish different modes in the spectrum, we introduce a normalized transition strength~\cite{chen2017collective}
\begin{equation}
    \Gamma_j(\hat{O})=\frac{\left| \left\langle j\left  |\hat{O}\right| 0\right\rangle \right|}{\text{max}\left[\left| \left\langle j\left  |\hat{O}\right| 0\right\rangle \right|^\infty_{j=1}\right]},
\end{equation}
with the ground state $|0\rangle$ and the $j$-th mode quasiparticle eigenfunction $|j\rangle = u_j+v_j$ in terms of the obtained quasiparticle amplitudes $u_j$ and $v_j$. Therefore, a transition strength $\Gamma_j(\hat{O})$ with a value close to 1 indicates the close relation between the specific quasiparticle excitation mode $j$ and the related perturbation operator $\hat{O}$. For example, we choose the operators $\hat{x}$ (or $\hat{y}$), $\hat{r}^2$, and $\hat{x}\hat{y}$ in this work to identify the associated dipole, breathing, and scissors modes, respectively. In addition, we can also distinguish between different excitation modes by substituting the frequency $\omega_j$ and amplitudes ($u_j$,$v_j$) of the quasiparticle into Eq.~\eqref{eq:linearized} and analyzing the time evolution of the density pattern.
\begin{figure*}[t]
\centering
\includegraphics[width=0.96\textwidth]{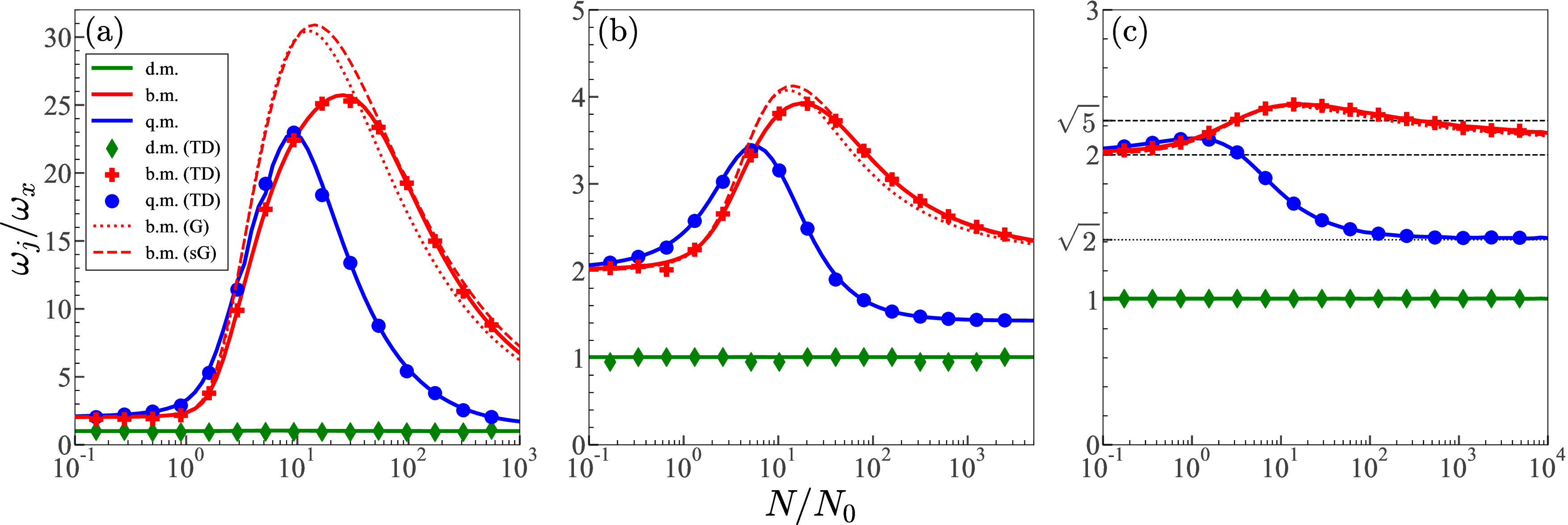}
\caption{Frequencies $\omega_j/\omega_x$ of specific collective modes as a function of the number of atoms $N/N_0$ at three values of the trapping strength (a) $\lambda_x=0.01$, (b) $\lambda_x=0.1$, and (c) $\lambda_x=1$, calculated by the linearization technique (solid lines), time-dependent extended GPE (symbols), a sum-rule approach based on Gaussian (dotted lines) and super-Gaussian (dashed lines) {\it Ansatz}, respectively. Here, the results of dipole, breathing, and quadrupole modes are denoted in green, red, and blue, respectively. The other parameters are the same as in Fig.~\ref{fig1}.}
\label{fig4}
\end{figure*}

To emphasize the specific collective modes and verify the accuracy of the calculated excitation frequency, we illustrate merely the results of the dipole, breathing, and quadrupole modes in Fig.~\ref{fig4}. It is worth noting that, the frequencies of the quadrupole and scissors modes in the current isotropic geometry are exactly degenerate (see the blue dots and the yellow crosses in Fig.~\ref{fig3}) and thus we analyze only the quadrupole mode in the following. In Fig.~\ref{fig4}, the corresponding frequencies are qualitatively and quantitatively compared by using the linearization technique, the time-dependent extended GPE, and the sum-rule approach, which are shown in solid lines, filled symbols, and dotted (dashed) lines, respectively. In general, three mode frequencies of the first two approaches show an excellent agreement in three subplots, which verifies the correctness of our numerical results to some extent. In contrast, owing to the inappropriate initial {\it Ansatz} for the condensate wave function, the analytic breathing-mode frequency obtained by the sum-rule approach may deviate from those of the first two approaches for intermediate-sized droplet, while the qualitative behaviors in the excitation frequency of all three approaches are consistent. 

For sufficiently weak trapping potential with $\lambda_x=0.01$ as shown in Fig.~\ref{fig4} (a), we anticipate to reproduce the results as in the uniform case~\cite{sturmer2021breathing}. Indeed, in the ideal gas limit, the frequency for both the breathing mode and the quadrupole mode is twice of the harmonic trapping frequency, $2\omega_x$. As the number of atoms $N$ increases, the excitation frequencies for the two modes start to diverge from each other, moving further apart. This illustrates a non-monotonic relationship between the number of atoms and excitation frequency for each mode. A distinct peak is observed for each mode at intermediate atomic numbers, with the excitation frequencies saturating at different values as $N$ increases. In the small- and large-droplet limits, i.e., $N/N_0\leq1$ and $N/N_0\geq100$, the droplet transitions from a Gaussian-like state to a flat-top quantum droplet state, which can be well captured by the Gaussian and super-Gaussian {\it Ansatz}, respectively~\cite{karlsson1992optical,sturmer2021breathing,tsoy2006dynamical,otajonov2019stationary,baizakov2011variational}. As a consequence, the analytic expression of the breathing-mode frequencies, i.e., Eq.~\eqref{omega_b_sG} and Eq.~\eqref{omega_b_G} denoted by the dashed and dotted lines, agrees approximately with the numerical results (in solid lines) in two limits. In the intermediate regime, however, both the Gaussian and super-Gaussian {\it Ansatz} adopted here can not effectively describe the true ground state and the analytic result deviates significantly from the numerical results by means of the linearization technique.

In Fig.~\ref{fig4} (b) and Fig.~\ref{fig4} (c), as the trapping potential is strengthened, we observe that the previously prominent peaks in the excitation frequencies become more subdued and smoothed out. The sum-rule approach, which fails to reproduce the breathing mode frequency for a moderate number of atoms in a weak potential, proves to be more effective when subjected to a strong confinement. Additionally, the peak turning points shift towards relatively small values of $N$, mirroring the trends of the minimum seen in the radius and chemical potential, as illustrated in Fig.~\ref{fig2} and discussed in the preceding section. Besides, in the large-$N$ limit, it's easy to see that the frequency of quadrupole mode saturates to be about $1.4\omega_x$, which is close to the one $\sqrt{2}\omega_x$ of a conventional two-dimensional Bose-Einstein condensates~\cite{ho1999quasi}, denoted by the horizontal grey dotted line. Meanwhile, the breathing-mode frequency tends towards a value within the range $[2,\sqrt{5}]\omega_x$, which are the analytic breathing-mode frequencies for two- and three-dimensional conventional weakly-interacting Bose-Einstein condensates respectively~\cite{stringari1996collective,ho1999quasi}, shown by two horizontal grey dashed lines. In all three subplots, the dipole mode frequency shown in green color is irrelevant to the number of atoms or interaction strength and takes the value of harmonic trapping frequency, i.e., $\omega_\mathrm{d.m.}=\omega_x$, consistent with the Kohn's theorem~\cite{Kohn1961cyclotron}.

%%%%%%%%%%%%%%%%%%%%%%%%%%%%%%%%%%%%%%%%%%%%%%%%%%%%%%
\subsection{Anisotropy in harmonic traps}

In this section, we will explore the impact of the essential anisotropy inherent in the external harmonic traps on the static density distributions and the frequencies of collective modes. In the calculations, the ratio parameter $\kappa\equiv\omega_y/\omega_x$ of harmonic trapping frequencies is employed to adjust the degree of anisotropy.

\begin{figure}[t]
\centering
\includegraphics[width=0.48\textwidth]{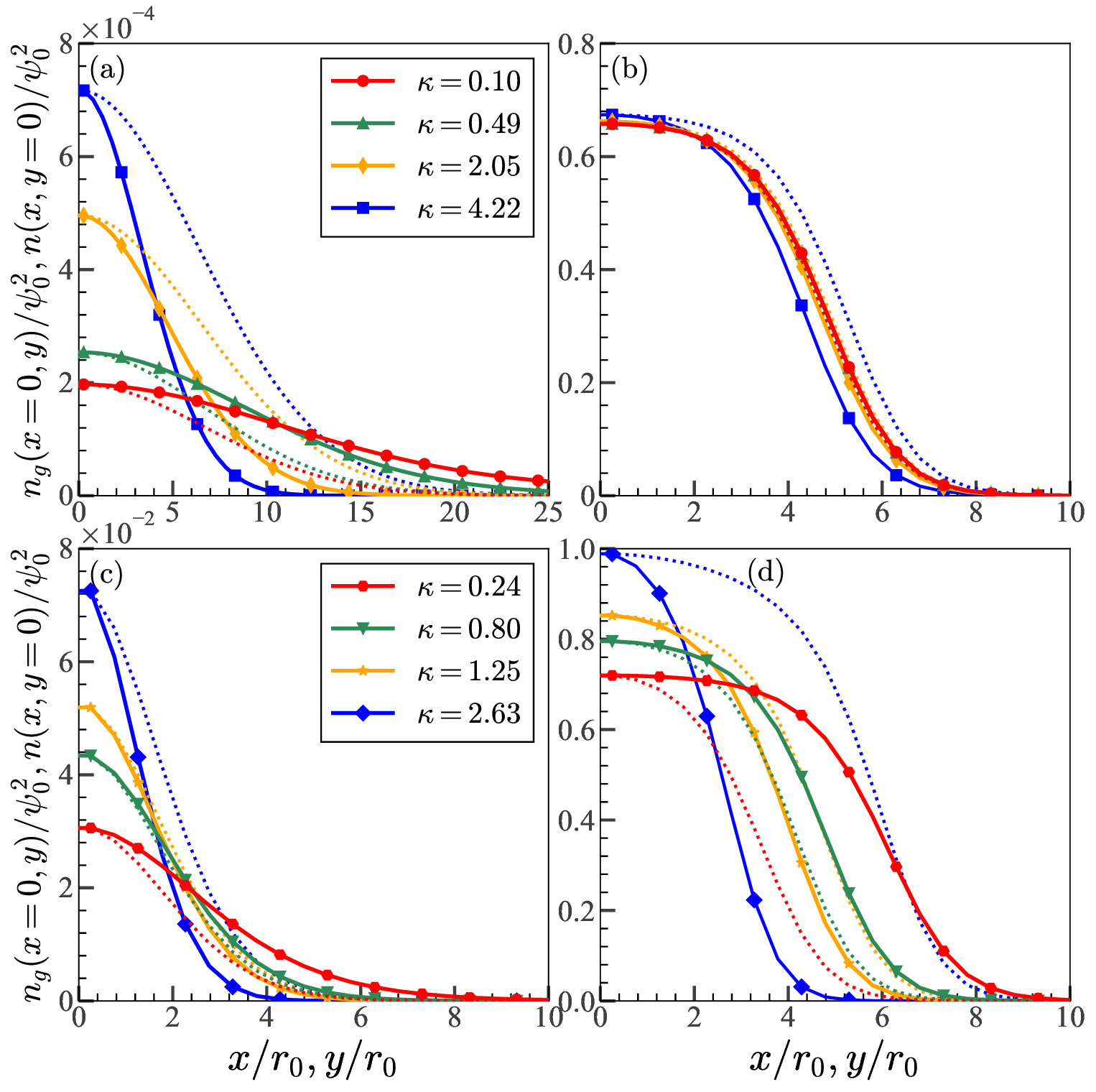}
\caption{Typical density profiles $n_g(0,y)$ (i.e., solid lines), $n_g(x,0)$ (i.e., dotted lines) of the droplet at four sets of the anisotropy parameter $\kappa\equiv\omega_y/\omega_x$. Here we take $\lambda_x=0.01$ for (a) $N/N_0=0.1$, (b) $N/N_0=50$, and $\lambda_x=0.1$ for (c) $N/N_0=1$, (d) $N/N_0=50$. The colored lines with dots, triangles, diamonds, and squares denote $\kappa=0.10$, 0.49, 2.05, and 4.22 in (a) and (b), and the colored lines with hexagons, inverted triangles, asterisks, and rhomboids denote $\kappa=0.24$, 0.80, 1.25, and 2.63 in (c) and (d), respectively. The other parameters are the same as in Fig.~\ref{fig1}.}
\label{fig5}
\end{figure}

We depict the cross-section density profiles $n_g(0,y)$ (solid lines) and $n_g(x,0)$ (dotted lines) of the droplet  in Fig.~\ref{fig5} at weak ($\lambda_x=0.01$) and moderate ($\lambda_x=0.1$) trapping strengths for various values of $\kappa$ and two scenarios of the number of atoms $N/N_0$. In Figs.~\ref{fig5} (a) and \ref{fig5} (b) with a very weak trapping potential $\lambda_x=0.01$, the density profile transitions from a conventional Gaussian-like shape at $N/N_0=0.1$ to a flat-top structure at $N/N_0=50$. A smaller droplet exhibits a Gaussian-like profile as shown in Fig.~\ref{fig5} (a), and even a slight anisotropy in the harmonic trap can readily result in density anisotropy. In sharp contrast, Fig.~\ref{fig5} (b) shows that a larger droplet with flat-top is stabilized by a significant self-bound effect via the LHY correction energy, which effectively mitigates the impact of anisotropy in the weak harmonic trap. As a result, a moderate range of anisotropy parameter $\kappa \sim 1$ has very little effect on these two cross-section densities, i.e., $n_g(0,y)$ and $n_g(x,0)$ remain the same and no obvious anisotropy appears. For a stronger trapping strength $\lambda_x=0.1$ in Fig.~\ref{fig5} (c) and \ref{fig5} (d), the droplet undergoes a transition from a Gaussian-like shape into a Thomas-Fermi-like distribution, which makes the anisotropic effect of harmonic traps more pronounced. As $\kappa$ increases, the confinement along the $y$-axis becomes progressively stronger compared to that along the $x$-axis, leading to a narrowing of the corresponding width. Thus, $\kappa$ with a value larger than unity implies a larger $x$-axis width and vise versa, giving rise to anisotropic density distributions of the droplet. These deviations between cross-section densities $n_g(0,y)$ and $n_g(x,0)$ can be seen clearly in the figure.

%and four values of $\kappa=0.24$, $0.80$, $1.25$, and $2.63$ for two cases of normalization $N/N_0 = 1$ and $50$. 
\begin{figure}[t]
\centering
\includegraphics[width=0.48\textwidth]{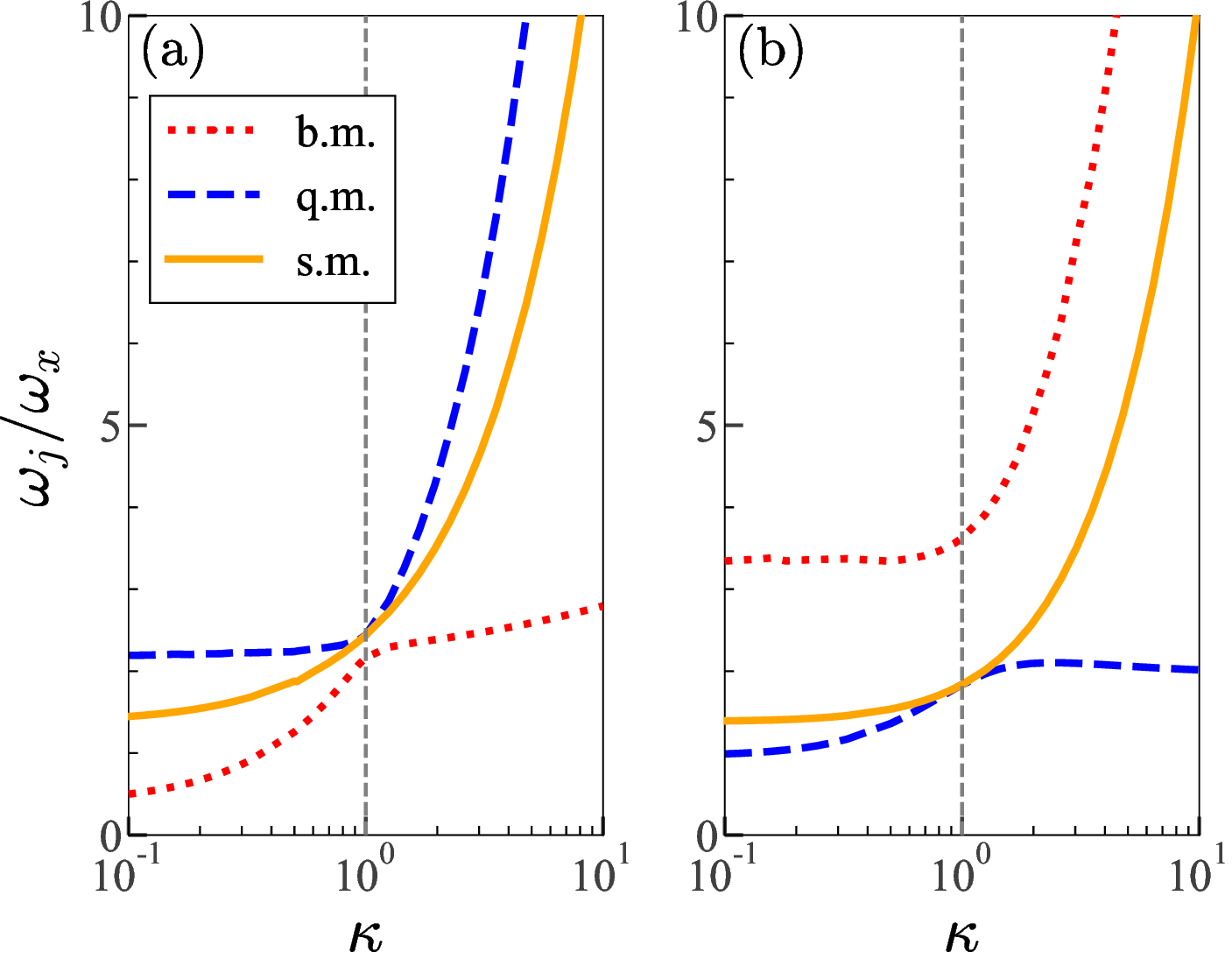}
\caption{Frequencies $\omega_j/\omega_x$ of breathing, quadrupole, and scissors modes (denoted by the dotted, dashed, and solid lines respectively), as a function of the anisotropy parameter $\kappa\equiv\omega_y/\omega_x$ at $\lambda_x=0.1$ for (a) $N/N_0=1$, and (b) $N/N_0=50$. The vertical dashed grey lines indicate the isotropic points, i.e., $\kappa=1$. The other parameters are the same as in Fig.~\ref{fig1}.}
\label{fig6}
\end{figure}

We further explore how the anisotrpy in the harmonic trap affect the excitation frequencies of three collective modes. We take two cases of number of atoms $N/N_0=1$ and $50$ at a moderate trapping potential $\lambda_x=0.1$ as an example and illustrate the excitation frequency as a function of the anisotropy parameter $\kappa$ in Fig.~\ref{fig6}. In general, the frequencies of these three modes show inconsistent dependence on $\kappa$, and the same mode frequency may exhibit diverse behaviors for small or large droplet. In both cases, however, there exists a frequency degeneracy between the quadrupole and scissors modes (i.e., blue-dashed and yellow-solid lines, respectively) at the isotropic point $\kappa=1$ denoted by the vertical dashed line, as mentioned in the last section. The degeneracy disappears immediately for arbitrarily small anisotropy in the trapping potential with $\kappa \neq 1$ and the frequencies of these two modes diverge notably from each other for $\kappa >1$ as well as for $\kappa <1$. In addition, in Fig.~\ref{fig6} (a) and ~\ref{fig6} (b), the frequencies of the breathing and quadrupole (or scissors) modes switch the upper and lower relative positions for $N/N_0=1$ and $50$, respectively. This is due to the breathing mode costs more energy in a large-sized droplet than that in a smaller one. The switched positions of mode frequencies are consistent with their $N$-dependent behavior in Fig.~\ref{fig4} (b) for isotropic case.

%%%%%%%%%%%%%%%%%%%%%%%%%%%%%%%%%%%%%%%%%%%%%%%%%%%%%%
%%%%%%%%%%%%%%%%%%%%%%%%%%%%%%%%%%%%%%%%%%%%%%%%%%%%%%
\section{CONCLUSIONS AND OUTLOOKS} \label{sec:conclusions}

In conclusion, we have studied the stationary properties and the collective excitations of a two-dimensional droplet formed from a two-component bosonic mixture in the presence of an external harmonic trap. The extended Gross-Pitaevskii equation (GPE) is derived from the energy functional, with which we numerically calculate the static density distribution, root-mean-square size of the droplet as well as the chemical potential. Our results in the limit of very weak trapping potential can well reproduce the predictions for self-bounded droplet in previous literature, such as the characteristic change in the droplet size and chemical potential at intermediate number of atoms in the droplet. We further explore the consequences of harmonic traps on the droplet and find interesting phenomena such as the clear softening of the emergent dip-like variation of radius with the number of atoms in the droplet for stronger confinement and the shifting of the minimum in the behavior of both radius and chemical potential as the number of atoms varies.

In addition, we have developed the linearization technique to investigate the low-energy collective excitations in trapped quantum droplet system. Several important collective modes such as the dipole, monopole, quadrupole, and scissors modes are intensively investigated in this work, and the calculated excitation frequencies are verified by utilizing the time-dependent extended GPE with designed perturbations to excite the corresponding modes. In particular, the frequencies of the monopole and breathing modes are further compared with the analytic predictions of a sum-rule approach based on the Gaussian and super-Gaussian variational approximation. We find that in weak trapping potentials, the mode frequencies exhibit typical non-monotonic behaviors with increasing droplet size, similar to the cases in one and three dimensional droplets. These variations in mode frequencies are strongly suppressed and smoothed out by relatively stronger trapping potentials, and the peaks tend to move to smaller values of $N$ as the droplet is trapped more tightly. 

Furthermore, we have considered the effect of anisotropy in harmonic traps on the density profiles and the excitation modes by introducing an anisotopy parameter $\kappa$. We find a moderate anisotropy in a weak trap does not significantly affect the density and leads to no anisotropy in density distribution, owing to the dominant self-bound effect supported by the LHY correction energy. Nonetheless, at sufficiently strong traps, the anisotropy in harmonic traps will easily shape the droplet in anisotropic density distribution. As for the collective modes, we find that the anisotropy breaks the degeneracy in the mode frequency between the quadrupole and scissors modes in an isotropic trap, and the frequencies of all collective modes deviate notably from each other as the anisotropy becomes stronger.

Further investigations might be carried out in the future to enhance the understanding of the ground-state characteristics and various collective behaviors of bose-mixture with imbalanced atomic masses under external potentials in low dimensions. For instance, the formation and quench of heteronuclear quantum droplets in one dimension is studied and the droplets can exhibit intriguingly distinct dynamic behaviors in free space or in an external harmonic confinement~\cite{minardi2019effective,mistakidis2021formation}. Besides, the asymmetric intra-atomic interaction strengths may also play a crucial role in the properties of quantum droplets~\cite{mithun2020modulational,parisi2020quantum,otajonov2022modulational}. It would be interesting to take into account these diverse parameters and explore the associated collective excitations and dynamic properties of quantum droplet in free space and in a trap.

\begin{acknowledgments}
This work is supported by the Natural Science Foundation of China (Grants No. 12074340 and No. 12204413), and the Science Foundation of Zhejiang Sci-Tech University (Grants No. 20062098-Y and No. 21062339-Y).
\end{acknowledgments}

\bibliography{quantumdroplets2d}

\end{document}